\documentclass[conference]{IEEEtran}
\usepackage{amsfonts}
\usepackage{amssymb}
\usepackage{graphicx}
\usepackage{amsmath}
\usepackage{epsfig}
\usepackage{setspace}
\usepackage{subfigure}
\usepackage{color}
\usepackage{bm}
\usepackage{cite}
\usepackage{authblk}

\newtheorem{definition}{Definition}
\newtheorem{lemma}{Lemma}
\newtheorem{proposition}{Proposition}

\newtheorem{rem}{Remark}

\begin{document}
\title{Optimal Distributed Beamforming for MISO Interference Channels}

\author[1]{Jiaming Qiu}
\author[2]{Rui Zhang}
\author[3]{Zhi-Quan Luo}
\author[1]{Shuguang Cui}
\affil[1]{ECE Department, Texas A\&M University, email: tristanqiu@tamu.edu, cui@ece.tamu.edu}
\affil[2]{ECE Department, National University of Singapore, email: elezhang@nus.edu.sg}
\affil[3]{ECE Department, University of Minnesota, email: luozq@ece.umn.edu}

 \maketitle

\begin{abstract}
We consider the problem of quantifying the Pareto optimal boundary in the achievable rate region over multiple-input single-output (MISO) interference channels, where the problem boils down to solving a sequence of convex feasibility problems after certain transformations. The feasibility problem is solved by two new distributed optimal beamforming algorithms, where the first one is to parallelize the computation based on the method of alternating projections, and the second one is to localize the computation based on the method of cyclic projections. Convergence proofs are established for both algorithms.
\end{abstract}

\section{Introduction}\label{introduction}
Traditional wireless mobile systems are designed with the cellular architecture, in which neighboring base stations (BSs) in different cells try to manage communications for their intended mobile stations (MSs) over non-overlapping channels. The associated inter-cell interferences, from non-neighboring cells, are treated as additive background noises. To improve the performance of traditional systems, most beyond-3G wireless technologies relax the frequency reuse constraint such that the whole frequency band becomes available for all cells. As such, joint signal processing across neighboring BSs is needed to cope with the strong inter-cell interferences in the future cellular systems.

In this paper, we study a particular type of multi-BS cooperation for downlink transmissions, where we assume a scenario with each BS equipped with multiple antennas and each MS equipped with a single antenna. Besides, only one MS is assumed to be active in each cell at any given time (over a particular frequency band). Our problem setup can be modeled as a multiple-input single-output (MISO) Gaussian interference channel (IC), termed as MISO-IC.

From an information-theoretic viewpoint, the best achievable rate region to date for an IC was established by Han and Kobayashi in \cite{HK}, termed as the Han-Kobayashi region, which utilizes rate splitting at transmitters, joint decoding at receivers, and time sharing among codebooks. The Han-Kobayashi region was simplified in \cite{HK_simplify} and proved to be within 1-bit of the capacity region of the Gaussian IC in \cite{One_bit}. However, in cellular systems, practical constraints often limit MSs to only implement single-user detection (SUD) schemes, i.e., treating the interference from all other unintended BSs as noise. Hence, in this work, we assume SUD at the MS receivers. With SUD, it has been shown that transmit beamforming is optimal for MISO IC in \cite{downlink} and \cite{X_Shang}. For the two-user case, Jorswieck \emph{et al.} \cite{complete} proved that the Pareto-optimal beamforming vectors can be represented as linear combinations of the zero-forcing (ZF) and maximum-ratio transmission (MRT) beamformers. Previous studies \cite{central1} and \cite{central2} over MISO-IC beamforming usually assumed a central processing unit with global knowledge of all the downlink channels, which may not be feasible in practical systems. To make the result more implementable, our work focuses on multi-cell cooperative downlink beamforming, which involves distributed computations based on the local channel knowledge at each BS. Such decentralized multi-cell cooperative beamforming problems were previously studied in \cite{Duality} based on the uplink-downlink duality to minimize the sum transmission power. Furthermore, a heuristic decentralized algorithm was developed in \cite{downlink} for multi-cell cooperative downlink beamforming based on the iterative updates of certain interference-temperature constraints across different pairs of BSs.

It has been discussed in \cite{downlink} that quantifying the Pareto optimal points in the achievable rate region over MISO IC may boil down to solving a sequence of convex feasibility problems after certain transformations, where the feasibility problems can be recast as second-order cone programming (SOCP) problems as shown in \cite{SOCP}. In this paper, we propose two algorithms to solve the resulting feasibility problem parallelly or distributively. In the first parallized beamforming algorithm based on alternating projections, we assume a computation-power limited centralized processing unit such that part of the computation duties need to be parallely conducted in each individual BS. In the second beamforming algorithm, localized sequential optimizations across the BSs are performed iteratively, where the need for a central processing unit is eliminated. Convergence in norm for both algorithms is established. Besides, a set of feasibility decision rules is established to implement our algorithms for practical engineering applications.

The rest of the paper is organized as follows. Section \ref{model} presents the MISO-IC model for multi-cell downlink beamforming, defines the Pareto optimality, and reviews the rate profile approach, which transforms the whole problem into solving a sequence of SOCP feasibility problems. Section \ref{PPB} proposes the parallelized algorithm to parallelly solve the SOCP feasibility problem based on the method of alternating projections. Section \ref{cyclic} presents the distributed beamforming algorithm to solve the SOCP feasibility problem based on the method of cyclic projections. Numerical examples are provided in Section \ref{simulations} with conclusions in Section \ref{conclusion}.

\emph{Notations}: Bold face letters, e.g., $\bm{x}$ and $\bm{X}$, denote vectors and matrices, respectively. $\bm{I}$ and $\bm{0}$ denote the identity matrix and the all-zero matrix, respectively, with appropriate dimensions. $diag(\bm{X}_1,\ldots,\bm{X}_n)$ defines a block diagonal matrix in which the diagonal elements are $\bm{X}_1,\ldots,\bm{X}_n$. $\left(  \cdot \right)^T$ and $\left(  \cdot \right)^H$ respectively denote the transpose and the Hermitian of  a matrix or a vector. $\mathbb{R}^{m \times n}$ and $\mathbb{C}^{m \times n}$ denote the space of $m \times n$ real matrices and the space of $m \times n$ complex matrices respectively. $\left\| \bm{x} \right\|$ denotes the Euclidean norm of a complex vector $\bm{x}$. All the $\log \left(  \cdot  \right)$ functions are with base 2 by default. ${\mathop{\rm Re}\nolimits} \left(  \cdot  \right)$ and ${\mathop{\rm Im}\nolimits} \left(  \cdot  \right)$ denote the real part and imaginary part of a complex argument respectively. $\left[ {{\bm{a}}_1 ; \ldots ;{\bm{a}}_n } \right]$ defines a vector that stacks ${\bm{a}}_1  \ldots {\bm{a}}_n$ into one column. By default, all the vectors are column vectors.

\section{System Model and Preliminaries}\label{model}

\subsection{Signal Model}
We address downlink transmissions in a cellular network consisting of $M$ cells, each having a multi-antenna BS to transmit an independent message to one active single-antenna MS. With the assumption that the same band is shared among all BSs for downlink transmissions, the system could be modeled as a $M$-user MISO-IC. Specifically, we assume that each BS is equipped with $K$ transmitting antennas, $K \geq 1$. With the assumption of single-user detection at each receiver, it has been shown in \cite{downlink} and \cite{X_Shang} that beamforming is optimal to maximize the rate region. Hence, the discrete-time baseband received signal of the active MS in the $i$th cell is given by
\begin{equation} \label{received-signal}
{y_i} = {\bm{h}}_{ii}^H{{\bm{\omega}}_i}{s_i} + \sum\limits_{j = 1,j \ne i}^M {{\bm{h}}_{ji}^H{{\bm{\omega }}_j}{s_j}}  + {z_i},~i = 1,2, \cdots ,M,
\end{equation}
where ${{\bm{\omega }}_i} \in {\mathbb{C}^{K}}$ denotes the beamforming vector at the $i$th BS; ${{\bm{h}}_{ii}} \in {\mathbb{C}^{K}}$ denotes the channel vector from the $i$th BS to its intended MS, while ${{\bm{h}}_{ji}} \in {\mathbb{C}^{K}}$ denotes the cross-link channel from the $j$th BS to the MS in the $i$th cell, $i \neq j$; $s_i$ denotes the symbol transmitted by the $i$th BS; and $z_i$ denotes the additive circular symmetric complex Gaussian (CSCG) noise at the $i$th receiver. It is assumed that ${z_i} \sim \mathcal{C}\mathcal{N}\left( {0,\sigma _i^2} \right)$ and ${z_i}$'s are independent.

We assume that the $i$th receiver only knows channel $\bm{h}_{ii}$, and decodes its own messages by treating interferences from all other BSs as noise. With SUD, the achievable rate for the $i$th MS is thus given as
\begin{equation}\label{rate}
{R_i} = \log \left( {1 + \frac{{{{\left| {{\bm{h}}_{ii}^H{\bm{\omega} _i}} \right|}^2}}}{{\sum\nolimits_{i \ne j} {{{\left| {{\bm{h}}_{ii}^H{\bm{\omega} _j}} \right|}^2} + \sigma _i^2} }}} \right),
\end{equation}
where the maximum transmission power is limited as
\begin{equation}
\left\| {{\bm{\omega }}_i } \right\|^2  \le P_i ,~i = 1,2, \ldots ,M,
\end{equation}
where $P_i$ is the power constraint at the $i$th BS.

\subsection{Pareto Optimality}
We define the achievable rate region for the MISO-IC to be the collection of rate-tuples for all MSs that can be simultaneously achievable under a certain set of transmit-power constraints:

\begin{align}
{\cal R}: = \bigcup\limits_{\{ {\bf{\omega }}_i \} :\left\| {{\bf{\omega }}_i } \right\|^2  \le P_i ,i = 1, \ldots M} {\left\{ \begin{array}{l}
 (r_1 , \ldots ,r_M ): \\
 0 \le r_i  \le R_i ({\bf{\omega }}_1 , \ldots ,{\bf{\omega }}_M ), \\
 i = 1, \ldots ,M \\
 \end{array} \right\}}.
\end{align}

The upper-right boundary of this region is called the Pareto boundary, since it consists of rate-tuples at which it is impossible to increase some user's rate without simultaneously decreasing the rate of at least one other users. To be more precise, the Pareto optimality of rate-tuple is defined as follows \cite{complete}.
\begin{definition}
A rate-tuple $\left( {{r_1} , \ldots ,{r_M} } \right)$ is Pareto optimal if there is no other rate-tuple $\left( {\hat r_1 , \ldots ,\hat r_M } \right)$ with $\left( {\hat r_1 , \ldots ,\hat r_M } \right) \\  \ge \left( {r_1 , \ldots ,r_M } \right)$ and $\left( {\hat r_1 , \ldots ,\hat r_M } \right)  \ne \left( {r_1 , \ldots ,r_M } \right)$, with the inequality being component-wise.
\end{definition}
In this paper, we are interested in searching the beamforming vectors for all BSs that lead to Parato optimal rate-tuples.

\subsection{Rate Profile Approach}
The rate profile approach \cite{rate_profile_literature} is an effective way to characterize the Pareto boundary of MISO-IC \cite{downlink}, where the key is that any rate tuple on the Pareto boundary can be obtained by solving the following optimization problem given a specified \emph{rate-profile vector}, $\bm{\alpha}  = \left( {{\alpha _1}, \ldots ,{\alpha _M}} \right)$:
\begin{eqnarray}\label{rate profile}
  \nonumber \mathop {\max }\limits_{{R_{sum}},\left\{ {{{\bm{\omega }}_i}} \right\}} && {R_{sum}}\\
  \nonumber s.t. && \log \left( {1 + \frac{{{{\left| {{\bm{h}}_{ii}^H{{\bm{\omega }}_i}} \right|}^2}}}{{{{\sum\nolimits_{i \ne j} {\left| {{\bm{h}}_{ji}^H{{\bm{\omega }}_j}} \right|} }^2} + \sigma _i^2}}} \right) \ge {\alpha _i}{R_{sum}},\\
  \nonumber &&~~~~i = 1,2, \ldots ,M,
 \\
  && {\left\| {{{\bm{\omega }}_j}} \right\|^2} \le {P_j},~j = 1,2, \ldots ,M,
\end{eqnarray}
where $\bm{\alpha}$ satisfies that ${\alpha _i} \ge 0,~1 \leq i \leq M$, and $\sum\nolimits_{i = 1}^M {{\alpha _i} = 1}$. Denote the optimal objective value of Problem (\ref{rate profile}) as $R_{sum}^*$. As shown in \cite{downlink}, $R_{sum}^* \cdot \bm{\alpha}$ corresponds to a particular Pareto optimal rate tuple. Hence, by exhausting all possible values for $\bm{\alpha}$, solving Problem (\ref{rate profile}) yields the whole Pareto boundary.

\subsection{SOCP Feasibility Problem}
Directly solving Problem (\ref{rate profile}) is usually difficult due to its non-convexity. However given the fact that the objective function is a single variable, we could adopt the bisection search algorithm to efficiently find $R_{sum}^*$ as shown in \cite{downlink}. Specifically, we could solve a sequence of the following feasibility problems each for a given $r_0$:
\begin{eqnarray}\label{feasibility}
  \nonumber \mathop {\max }\limits_{\left\{ {{{\bm{\omega }}_i}} \right\}} && 0 \\
  \nonumber s.t. && \log \left( {1 + \frac{{{{\left| {{\bm{h}}_{ii}^H{{\bm{\omega }}_i}} \right|}^2}}}{{{{\sum\nolimits_{i \ne j} {\left| {{\bm{h}}_{ji}^H{{\bm{\omega }}_j}} \right|} }^2} + \sigma_i^2}}} \right) \ge {\alpha _i}{r_0}, \\
  \nonumber && ~~~~i = 1,2, \ldots ,M,
 \\
   &&  {\left\| {{{\bm{\omega }}_j}} \right\|^2} \le {P_j},~j = 1,2, \ldots ,M.
\end{eqnarray}
Therefore if the above problem is feasible for $r_0$, it follows that $R_{sum}^* \geq {r_0}$; otherwise, $R_{sum}^* < {r_0}$. Hence, a bisection search over $R_{sum}$ can be done. However, Problem (\ref{feasibility}) is still non-convex.

As shown in \cite{SOCP}, we can adjust the phase of $\bm{\omega}_i$ in (\ref{feasibility}) to make ${\bm{h}}_{ii}^H{\bm{\omega }_i}$ real and non-negative without affecting the value of $\left| {\bm{h}_{ii}^H{\bm{\omega }_i}} \right|$. Hence, by denoting ${\beta _i} = {e^{{\alpha _i}{r_0}}} - 1, ~ i = 1,2, \ldots ,M$, Problem (\ref{feasibility}) can be recast as
\begin{align}\label{1recast}
  \nonumber \mathop {\max }\limits_{\left\{ {{{\bm{\omega }}_i}} \right\}} &~~~ 0 \\
  \nonumber s.t. &~~~ {\left( {{\bm{h}}_{ii}^H{{\bm{\omega }}_i}} \right)^2} \ge {\beta _i}\left( {{{\sum\nolimits_{i \ne j} {\left| {{\bm{h}}_{ji}^H{{\bm{\omega }}_j}} \right|} }^2} + \sigma _i^2} \right),\\
  \nonumber &~~~~~~ i=1,2, \ldots, M,\\
  \nonumber &~~~  {\bm{h}}_{ii}^H{\bm{\omega }_i} \geq 0,~i=1,2, \ldots , M,\\
  &~~~  {\left\| {{{\bm{\omega }}_j}} \right\|} \le \sqrt{P_j},~j = 1,2, \ldots ,M.
\end{align}
We further define $\bm{x} = {\left[ {{\bm{\omega }}_1} ; {{\bm{\omega }}_2} ;  \cdots ; {{\bm{\omega }}_M} ; 0 \right]}$, ${\bm{n}}_i  = \left[ {0; 0; \ldots ;0;\sigma _i } \right]$, ${\bm{S}_i} = \left[ {\begin{array}{*{20}{c}}
    \cdots  & {{I_K}} &  \cdots  & 0  \\
\end{array}} \right]$ with ${{\bm{S}}_i}{\bm{x}} = \bm{\omega} _i,~i = 1,2, \ldots ,M$, and ${\bm{A}}_i  = diag\left( {{\bm{h}}_{1i}^H ,{\bm{h}}_{2i}^H , \ldots ,{\bm{h}}_{Mi}^H ,0} \right)$. For convenience, we add a term ${\beta _i}{\left( {{\bm{h}}_{ii}^H{{\bm{\omega }}_i}} \right)^2}$ to both sides of the first constraint in Problem (\ref{1recast}) as
\begin{align}
\left( {1 + {\beta _i}} \right){\left( {{\bm{h}}_{ii}^H{{\bm{\omega }}_i}} \right)^2} \ge {\beta _i}\left( {\sum\nolimits_{j = 1}^M {{{\left| {{\bm{h}}_{ji}^H{{\bm{\omega }}_j}} \right|}^2}}  + \sigma _i^2} \right),
\end{align}
where $i = 1, \ldots ,M$. Finally, with our newly defined variables and coefficients, we recast Problem (\ref{1recast}) as
\begin{align}\label{SOCP feasible}
  \nonumber \mathop {\max }\limits_{\bm{x}}  &~~~ 0 \\
  \nonumber s.t. & ~~~{\sqrt {\beta _i} \left\| {{\bm{A}_i}{\bm{x}} + {{\bm{n}}_i}} \right\|} \le \sqrt {1 + {\beta _i}} \left( {{\bm{h}}_{ii}^H{{\bm{S}}_i}{\bm{x}}} \right), \\
   \nonumber &~~~~~~i = 1,2,\ldots,M,\\
   \nonumber & ~~~{{\bm{p}}^T}{\bm{x}} = 0, \\
   &~~~ \left\| {{{\bm{S}}_j}{\bm{x}}} \right\| \le \sqrt{{P_j}},~j = 1, \ldots ,M,
\end{align}
where vector $\bm{p}$ is of the same dimension as $\bm{x}$ with all zero elements except for the last one being 1, such that the last element of $\bm{x}$ is guaranteed to be 0.

Consequently, Problem (\ref{SOCP feasible}) is a SOCP problem, which can be efficiently solved by numerical tools \cite{ConvexOptimization}. However, directly solving Problem (\ref{SOCP feasible}) requires a centralized algorithm running at a control center, which may not be desired in certain engineering applications. Accordingly, there are usually two motivations for seeking distributed algorithms: one is to decompose the computations into multiple sub-programs such that the requirement for the central processing power is reduced; and the other is to localize computations such that no central control facility is required. In Section \ref{PPB} and Section \ref{cyclic}, we propose two algorithms based upon the above two motivations, respectively.

\section{Alternating Projections Based Distributed Beamforming}\label{PPB}
In this section, in order to reduce the requirement on processing power at the control center, we develop a downlink beamforming algorithm, termed as alternating projections based distributed beamforming (APB), to solve Problem (\ref{SOCP feasible}) parallelly in $M$ sub-problems. With our algorithm, the only processing power needed at the central unit is to calculate an average value over all the localized solutions from the $M$ BSs. The algorithm is iterative, where parallel optimizations across BSs are performed at each round. The convergence issue of APB is also studied in this section.

\subsection{APB Algorithm}\label{PP_A}
\begin{figure}[htbp]
    \begin{center}
    \includegraphics[width=3.2in]{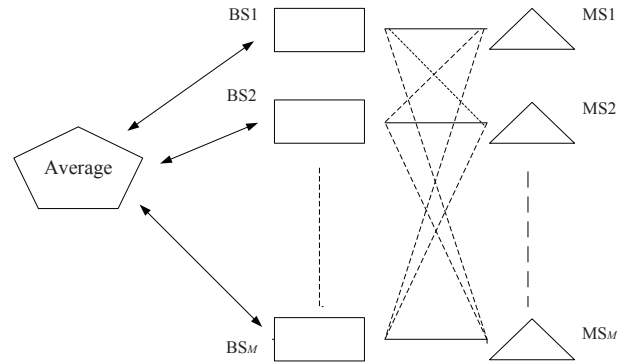}
    \caption{APB Scheme}
    \label{pp_structure}
    \end{center}
\end{figure}

At the initialization stage, the computation-limited centralized unit is assigned with the values for $M$, $K$, and $P_1 , \ldots ,P_M$. Then the central unit broadcasts the information to all BSs with an arbitrary initial point ${\bm{\tilde x}}_0  \in \mathbb{C}^{KM + 1}$.  It is assumed that the $i$th BS has the perfect knowledge of the channels from all BSs to the $i$th MS, i.e., all $\bm{h}_{ij}$'s. Furthermore, all BSs operate according to the same protocol described as follows. At the $n$th round, we denote the solution vector that the central unit broadcasts as ${\bm{\tilde x}}_{n-1}$. Then at the $i$th BS, the corresponding problem is expressed as
\begin{eqnarray}\label{subproblem avg}
  \nonumber \mathop {\min }\limits_{\bm{x}} && {\left\| {{\bm{x}} - {{{\bm{\tilde x}}}_{n - 1}}} \right\|} \\
  \nonumber s.t. && \sqrt {{\beta _i}} {\left\| {{{\bm{A}}_i}{\bm{x}} + {{\bm{n}}_i}} \right\|} \le \sqrt {1 + {\beta _i}} ({\bm{h}}_{ii}^H{{\bm{S}}_i}{\bm{x}}), \\
  \nonumber && {{\bm{p}}^T}{\bm{x}} = 0, \\
   && {\left\| {{{\bm{S}}_j}{\bm{x}}} \right\|} \le \sqrt {{P_j}} ,j = 1,2, \ldots ,M,
\end{eqnarray}
where ${{{\bm{\tilde x}}}_{n - 1}} = \frac{1}{M}\sum\nolimits_{i = 1}^M {{\bm{x}}_{n - 1}^{(i)}}$, with ${\bm{x}}_{n-1}^{(i)}$ denoting the optimal solution for Problem (\ref{subproblem avg}) of the ($n-1$)th round at the $i$th BS. A rough description of the algorithm is depicted in Fig. \ref{pp_structure}.

\begin{rem}\label{rem1}
Note that if Problem (\ref{subproblem avg}) is infeasible at the $k$th BS ($k \in \left\{ {1, \ldots ,M} \right\}$), we can directly claim that the associated Problem (\ref{SOCP feasible}) is infeasible and quit APB. As such, from now on we only focus on the cases where Problem (\ref{subproblem avg}) is always feasible at each individual BS, and run APB to check when the overall problem in (\ref{SOCP feasible}) is feasible and when it is not. With a feasible Problem (\ref{subproblem avg}), we need the optimal solution ${\bm{x}_n^{(i)}}$ to satisfy all the transmitter power constraints and the $i$th receiver's SNR demand. The reason why we keep all $M$ power constraints at each individual BS is for that fast convergence, which can be observed from simulations. Since all the $P_j$ values are typically predetermined in cellular systems, no extra system overhead is needed. In the second-order cone constraint of Problem (\ref{subproblem avg}), directly using the term ${\bm{h}}_{ii}^H{\bm{S}_i}\bm{x}$ implies that ${\mathop{\rm Im}\nolimits} \left( {{\bm{h}}_{ii}^H{\bm{S}_i}\bm{x}} \right) = 0$ and ${\bm{h}}_{ii}^H{\bm{S}_i}\bm{x} \ge 0$.
\end{rem}

\subsection{Convergence Analysis}
Since APB is iterative, the convergence is an important issue to address. The convergence of APB is formally stated as follows.

\begin{proposition}\label{convergence avg}
As $n$ increases, the optimal solution ${\bm{x}_n^{(i)}}$ for Problem (\ref{subproblem avg}) converges in norm to the limit ${\bm{\tilde x}^i}$ when Problem (\ref{SOCP feasible}) is either feasible or infeasible. Furthermore, the averaged solution ${\bm{\tilde x}_n}$ also converges in norm to ${\bm{\hat x}^*}$ satisfying that $\frac{1}{M}\sum\nolimits_{i = 1}^M {{\bm{\tilde x}^i}}  = {\bm{\hat x}^*}$. In particular, if Problem (\ref{SOCP feasible}) is feasible, all ${\bm{\tilde x}^i}$'s coincide in the same point $\bm{\tilde x}$ that lies in the feasible set of Problem (\ref{SOCP feasible}) with ${\bm{\tilde x}} = {\bm{\hat x}^*}$. If Problem (\ref{SOCP feasible}) is infeasible, ${\bm{\tilde x}^i}$'s do not coincide in the same solution.
\end{proposition}

\emph{Proof:}
1) For the case of Problem (\ref{SOCP feasible}) being feasible, we have the following proof.

We first introduce the concept of finding the closest point to some given point  in a closed convex set and alternating projections.

In mathematics, a Hilbert space $H$ is defined with the inner product $\left\langle {\bm{x},\bm{y}} \right\rangle$ and the induced norm $\left\| \bm{x} \right\| = \sqrt {\left\langle {\bm{x},\bm{x}} \right\rangle } $. If $S$ is a nonempty closed convex set in $H$, Riesz \cite{Riesz} states that each $\bm{x} \in H$ has a unique best approximation (or nearest point) ${P_S}(\bm{x})$ in $S$. That is, $\left\| {\bm{x} - {P_S}(\bm{x})} \right\| < \left\| {\bm{x} - \bm{y}} \right\|,~\forall y \in S\backslash \{ {P_S}(\bm{x})\}$. The mapping ${P_S}:H \to S$ is called the \emph{projection} onto $S$, i.e., finding the closest point to $\bm{x}$ in a closed nonempty convex set. In this paper, we use the general Euclidean inner product definitions $\left\langle {\bm{x},\bm{y}} \right\rangle  = {\bm{x}^H}\bm{y}$ in the complex space and $\left\langle {\bm{x},\bm{y}} \right\rangle  = {\bm{x}^T}\bm{y}$ in the real space.

\begin{definition}
Suppose $C_1$ and $C_2$ are two closed nonempty convex sets in $H$ with corresponding projections $P_1$ and $P_2$. Let $C = C_1  \cap C_2$ and fix a starting point ${x_0} \in H$. Then the sequence of \emph{alternating projections} is generated by
\begin{align*}
x_1  = P_1 x_0 ,x_2  = P_2 x_1 , x_3=P_1 x_2, \ldots , \\ x_N  = P_2 x_{N - 1} ,x_{N + 1}  = P_1 x_N , \ldots
\end{align*}
\end{definition}

Let $F_i$ denote the feasible set of Problem (\ref{subproblem avg}) at the $i$th BS, ${F_i} \ne \emptyset$, and $F = \bigcap\nolimits_{i = 1}^M {{F_i}} \neq \emptyset$; note that $F$ is exactly the feasible set of Problem (\ref{SOCP feasible}). Thus solving Problem (\ref{subproblem avg}) at the $i$th BS can be viewed as finding the \emph{closest} point to ${\bm{\tilde x}_{n - 1}}$ in a non-empty closed convex set $F_i$, i.e., the projection of ${\bm{\tilde x}_{n - 1}}$ onto $F_i$. Next we transform the variable defined over the complex Hilbert space to a double-dimensioned real Hilbert space such that we can use some existed results in alternating projections. We transform ${\bm{x}} \in {\mathbb{C}^{N'}}$ to ${\bm{\bar x}} \in {\mathbb{R}^{2N'}}$ by letting ${\bm{\bar x}} = \left[ {{\mathop{\rm Re}\nolimits} ({\bm{x}});{\mathop{\rm Im}\nolimits} ({\bm{x}})} \right]$, where $N'=KM+1$. Similarly, we map the complex set $F_i$ to a double-dimensioned real set $F'_i$, and map ${{{\bm{\tilde x}}}_{n - 1}}$ to a double-dimensioned real vector ${{{\bm{\hat x}}}_{n - 1}} = \left[ {{\mathop{\rm Re}\nolimits} ({{{{\bm{\tilde x}}}_{n - 1}}});{\mathop{\rm Im}\nolimits} ({{{\bm{\tilde x}}}_{n - 1}})} \right]$. We rewrite Problem (\ref{subproblem avg}) as
\begin{eqnarray}\label{real avg}
  \nonumber \mathop {\min }\limits_{{\bm{\bar x}} \in {\mathbb{R}^{2N'}}}  && {\left\| {{\bm{\bar x}} - {{{\bm{\hat x}}}_{n - 1}}} \right\|} \\
  \nonumber s.t. && \sqrt {{\beta _i}} {\left\| {{{\pmb{\bar A}}}_i}{\bm{\bar x}} + {{{\bm{\bar n}}}_i} \right\|} \le \sqrt {1 + {\beta _i}} \left( {{\bm{\bar h}}_{ii}^H{{{\bm{\bar S}}}_i}{\bm{\bar x}}} \right), \\
  \nonumber  && {\bm{d}}_i^H {\bm{\bar x}} = 0, \\
  \nonumber  && {{{\bm{\bar p}}}^T}{\bm{\bar x}} = 0, \\
   && {\left\| {{{{\bm{\bar S}}}_j}{\bm{\bar x}}} \right\|} \le \sqrt {{P_j}} ,~j = 1, \ldots ,M,
\end{eqnarray}
where \[{\bm{{\bar A}}_i} = \left[ {\begin{array}{*{20}{c}}
   {{\mathop{\rm Re}\nolimits} \left( {{\bm{A}_i}} \right)} & { - {\mathop{\rm Im}\nolimits} \left( {{\bm{A}_i}} \right)}  \\
   {{\mathop{\rm Im}\nolimits} \left( {{\bm{A}_i}} \right)} & {{\mathop{\rm Re}\nolimits} \left( {{\bm{A}_i}} \right)}  \\
\end{array}} \right],~{\bm{\bar h}}_{ii}^H = \left[ {\begin{array}{*{20}{c}}
   {{\mathop{\rm Re}\nolimits} \left( {{\bm{h}}_{ii}^H} \right)} & { - {\mathop{\rm Im}\nolimits} \left( {{\bm{h}}_{ii}^H} \right)}  \\
\end{array}} \right],\]

\[{{{\bm{\bar S}}}_i} = \left[ {\begin{array}{*{20}{c}}
   {{{\bm{S}}_i}} & {{{\bm{S}}_i}}  \\
\end{array}} \right],~{\bm{d}}_{i}^H = \left[ {\begin{array}{*{20}{c}}
   {{\mathop{\rm Im}\nolimits} \left( {{\bm{h}}_{ii}^H} \right)} & {{\mathop{\rm Re}\nolimits} \left( {{\bm{h}}_{ii}^H} \right)}  \\
\end{array}} \right]\left[ {\begin{array}{*{20}{c}}
   {{{\bm{S}}_i}} & {{{\bm{S}}_i}}  \\
\end{array}} \right],\]

\[
{\bm{\bar n}}_i  = \left[ {{\bm{n}}_i ;{\bm{0}}} \right],{\bm{\bar p}} = \left[ {{\bm{p}};{\bm{p}}} \right].
\]

From the constraints of Problem (\ref{real avg}), we observe that the feasible set $F'_i$ is the intersection of a collection of second-order cones, some subspaces, and some norm balls, which is nonempty closed and bounded. Next we show how to transform our algorithm into a problem of alternating projections. Let's define two product sets:
\begin{equation}
\nonumber T:{F'_1} \times {F'_2} \times  \cdots {F'_M},
\end{equation}
and
\begin{equation*}
U:~\left\{ {\left( {{\bm{a}},{\bm{a}}, \ldots ,{\bm{a}}} \right):{\bm{a}} \in \mathbb{R}^{2N'} } \right\}
\end{equation*}
Meanwhile£¬ we define two new variables ${\bm{x}_k},{\bm{y}_k} \in {\mathbb{R}^{2N'M}}$ as
\begin{equation}
{{\bm{x}}_k} = \left[ {{\bm{\bar x}}_k^{(1)};{\bm{\bar x}}_k^{(2)}; \cdots ;{\bm{\bar x}}_k^{(M)}} \right],~{{\bm{y}}_k} = \left[ {{{{\bm{\hat x}}}_k};{{{\bm{\hat x}}}_k}; \cdots ;{{{\bm{\hat x}}}_k}} \right].
\end{equation}
Obviously, ${\bm{x}_k} \in T,~{\bm{y}_k} \in U$. By the results of Pierra in \cite{Pierra}, we have the following two lemmas:
\begin{lemma}
Solving Problems (\ref{subproblem avg}) for $i=1, \ldots, M$ in parallel at the $k$th round is equivalent to projecting vector $\bm{y}_{k-1}$ onto the closed convex set $T$ and obtaining $\bm{x}_k$.
\end{lemma}

\begin{lemma}
Computing $\frac{1}{M}\sum\limits_{i = 1}^M {{\bm{x}}_k^{(i)}}$ is equivalent to projecting $\bm{x}_k$ onto $U$ and getting $\bm{y}_k$.
\end{lemma}

Therefore, APB can be interpreted as alternating projections between $T$ and $U$. Note that the idea of \emph{Alternating Projections} was first proposed by von Neumann in \cite{Neumann}, where only subspaces are assumed as the projection sets. Then many researchers extended this technique to more general scenarios \cite{Cheney}, \cite{Alternating}. For alternating projections between two non-empty closed convex sets $C_1$ and $C_2$, Cheney \cite{Cheney} proved that convergence in norm is always assured when either (a) one set is compact, or (b) one set is of finite dimension. Since set $T$ is bounded and our underlying Hilbert space is of finite dimension, both conditions (a) and (b) are satisfied. Therefore, APB always leads to strong convergence, i.e., convergence in norm, due to the facts that the numbers of cells and antennas are always finite. As shown in \cite{Alternating}, all ${\bm{\tilde x}^i}$'s will coincide into the same point $\bm{\tilde x}$ that lies in $F$.

2) For the case of Problem (\ref{SOCP feasible}) being infeasible, we have the following proof.

With $F = \emptyset$, the convergence of APB is still equivalent to the convergence of alternating projections between $T$ and $U$, where Cheney's results in \cite{Cheney} are applicable in this case. Thus, the convergence in norm is still valid for infeasible cases. Besides, it is easy to verify that $\frac{1}{M}\sum\nolimits_{i = 1}^M {{{{\bm{\tilde x}}}^i}}  = {{{\bm{\hat x}}}^*}$. However, ${\bm{\tilde x}^i}$'s do not coincide into the same point.

We now complete the proof for Proposition \ref{convergence avg}.~$\Box$

\subsection{Practical Feasibility Decision Rules}\label{dec rules}
With convergence in norm for APB established, we now need to establish some practical feasibility check rules to correctly terminate APB when it converges.

From Lemma 1 and Lemma 2, we know that the feasibility of Problem (\ref{SOCP feasible}) is totally determined by whether $T$ and $U$ intersect or not. By Proposition 1, if Problem (\ref{SOCP feasible}) is feasible, all convergent solutions ${{\bm{\tilde x}}^1}, \ldots ,{{\bm{\tilde x}}^M},{{\bm{\hat x}}^*}$ coincide at a common point $\bm{x}^*$ which belongs to $F$. In this case, all optimal values of Problems (\ref{subproblem avg}) converge to 0. On the other hand, if any of the optimal values of Problems (\ref{subproblem avg}) do not converge to 0, Problem (\ref{SOCP feasible}) is infeasible. Based on the above discussions, we develop the following APB terminating procedures:\\
\textbf{Step 1}: We set two threshold parameters $\epsilon$ and $\xi$. The selection of $\epsilon$ and $\xi$ affects the effectiveness of the algorithm.\\
\textbf{Step 2}: Initialization: Let ${v_i},~1\leq i \leq M,$ be the optimal value of Problem (\ref{subproblem avg}) at the $i$th cell in the current computation round, $v_i^*,~1\leq i \leq M,$ be the optimal value Problem (\ref{subproblem avg}) at the $i$th cell in the previous computation round, and ${\rm{flag}}[i]$,~$1 \leq i \leq M$ be the flags for the $M$ BSs. At the beginning, we set ${v_1}, \ldots ,{v_M}$ and ${{\rm{flag}}[1]}, \ldots, {{\rm{flag}}[M]}$ all zeros.\\
\textbf{Step 3}: Repeat: For $i = 1, \ldots , M$, the $i$th BS solves Problem (\ref{subproblem avg}) and compares $v_i$ against $v_i^*$. If $\left| {{v_i} - v_i^*} \right| \ge \epsilon$, we refresh ${v_i}:~{v_i} = v_i^*$ and proceed to Step 4; if $\left| {{v_i} - v_i^*} \right| < \epsilon$, we compare $v_i^*$ with $\xi$: If $v_i^* > \xi$, we claim that Problem (\ref{SOCP feasible}) is infeasible and stop; otherwise, we mark this cell as ${\rm{flag}}[i] = 1$ and proceed to Step 4.\\
\textbf{Step 4}: If ${\rm{flag[}}i] = 1$ for all $i = 1, \ldots ,M$, we claim that the Problem (\ref{SOCP feasible}) is feasible, then stop. Otherwise, return to Step 3.

\begin{rem}
Note that here we applied several approximations in making the decisions. First, we claim that Problem (\ref{subproblem avg}) at the $i$th BS converges when $\left| {v_i} - v_i^* \right| < \epsilon$. Thus, $v_i^*$ is considered as the limit of $i$th BS's optimal solution. Second, we set $\xi$ as the threshold dividing zero and non-zero values: If $v_i^* > \xi$, we consider the limit non-zero, and vice versa. In simulations, we usually set both $\epsilon$ and $\xi$ small with $\xi \gg \epsilon$. For example, $\epsilon = 0.002$ and $\xi=0.1$ are chosen for the simulation results in Section \ref{simulations}.
\end{rem}

\section{Cyclic Projections Based Distributed Beamforming}\label{cyclic}
In this section, to localize computations such that no central control unit is required, we propose a decentralized algorithm that practically implements the multi-cell cooperative downlink beamforming. It is still assumed that the $i$th BS in the cellular network has the perfect knowledge of the channels from all BSs to the $i$th MS. Similar to APB, we decompose Problem (\ref{SOCP feasible}) to $M$ sub-problems and compute them at $M$ BSs individually. In particular, the $M$ problems are solved sequentially at each round, and the algorithm proceeds iteratively, which is termed as Cyclic Projections Based Distributed Beamforming (CPB).

\subsection{CPB Algorithm}
\begin{figure}[htbp]
    \begin{center}
    \includegraphics[width=3.2in]{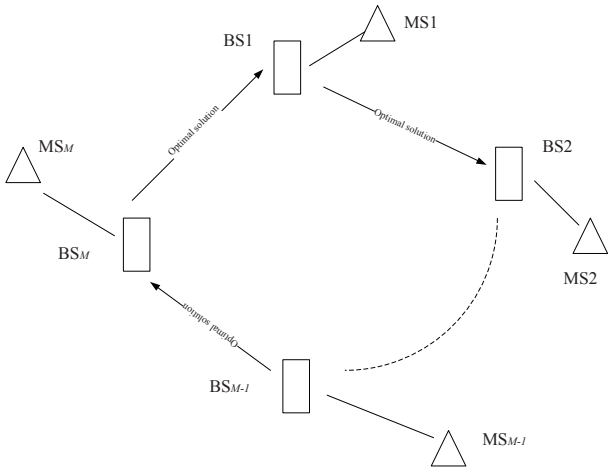}
    \caption{CPB Scheme}
    \label{cp_structure}
    \end{center}
\end{figure}

A certain cyclic update order among the $M$ BSs needs to be determined at the initialization stage, where the 1st BS sends its solution to the 2nd, $\ldots$, the ($M-1$)th BS sends its solution to the $M$th BS, and the $M$th BS sends its solution to the 1st, in a cyclic fashion. At the beginning, the $M$ BSs should obtain the values for $M$, $K$, and ${P_1},\ldots,{P_M}$. The algorithm starts from the 1st BS, after choosing an arbitrary initial point $\bm{x}_0$, it solves the following problem
\begin{eqnarray}\label{subproblem cyclic initial}
  \nonumber \mathop {\min }\limits_{{\bm{x}} \in {\mathbb{C}^{KM + 1}}} && {\left\| {{\bm{x}} - {\bm{x}}_0} \right\|} \\
  \nonumber s.t. && \sqrt {{\beta _1}} {\left\| {{{\bm{A}}_1}{\bm{x}} + {{\bm{n}}_1}} \right\|} \le \sqrt {1 + {\beta _1}} ({\bm{h}}_{11}^H{{\bm{S}}_1}{\bm{x}}), \\
  \nonumber && {{\bm{p}}^T}{\bm{x}} = 0, \\
  && {\left\| {{{\bm{S}}_j}{\bm{x}}} \right\|} \le \sqrt {{P_j}} ,~j = 1,2, \ldots ,M,
\end{eqnarray}
where the optimal solution for the above problem is labelled as ${\bm{x}}_1^{(1)}$ and sent to the 2nd BS. Then the other BSs begin to solve their own problems sequentially according to the predefined order. In particular, at the $n$th round the $i$th BS ($i\geq2$) solves the following problem
\begin{eqnarray}\label{subproblem cyclic}
  \nonumber \mathop {\min }\limits_{{\bm{x}} \in {\mathbb{C}^{KM + 1}}} && {\left\| {{\bm{x}} - {\bm{x}}_n^{(i - 1)}} \right\|} \\
  \nonumber s.t. && \sqrt {{\beta _i}} {\left\| {{{\bm{A}}_i}{\bm{x}} + {{\bm{n}}_i}} \right\|} \le \sqrt {1 + {\beta _i}} ({\bm{h}}_{ii}^H{{\bm{S}}_i}{\bm{x}}), \\
  \nonumber && {{\bm{p}}^T}{\bm{x}} = 0, \\
  && {\left\| {{{\bm{S}}_j}{\bm{x}}} \right\|} \le \sqrt {{P_j}} ,~j = 1,2, \ldots ,M,
\end{eqnarray}
where ${\bm{x}}_n^{(i-1)}$ is the solution sent over by the preceding BS, and ${\bm{x}}_n^{(i)}$ is used to denote the newly solved optimal solution. For simplicity, we refer the problem in (\ref{subproblem cyclic}) as a cyclic subproblem. Such a scheme is illustrated in Fig. \ref{cp_structure}.

\begin{rem}
Obviously, the constraints in Problem (\ref{subproblem cyclic}) and Problem (\ref{subproblem avg}) are the same. Therefore, we have the similar discussions as in Remark \ref{rem1}: We assume that all the cyclic subproblems are feasible when CPB is executed; otherwise, we directly claim that Problem (\ref{SOCP feasible}) is infeasible.
\end{rem}

\subsection{Convergence Analysis}
We first introduce the concept of cyclic projections.
\begin{definition}
Suppose $C_1,C_2, \ldots,C_r$ are closed convex sets in the Hilbert space $H$ with $C =  \cap _1^r C_i$, and let $P_i$ be the projection for $C_i$, $i=1,2,\ldots,r$. The operation of \emph{cyclic projections} is an iterative process that can be described as follows. Start with any point $x \in H$, and define the sequence ($x_n$) $(n = 1,2, \ldots)$ by
\begin{align}
x_0  = x, &x_1 = P_1(x_0),\ldots,~\text{and}~x_n  = P_{n\bmod r} \left( {x_{n - 1} } \right),
\end{align}
where $P_k(.)$ is the projection operator to $C_k$.
\end{definition}

In the literature, Bregman \cite{Bregman} showed that the above sequence generated by cyclic projections always converges weakly to some point $W_C \left( x \right) \in C$ provided that $C \neq \emptyset$, and Gubin \cite{Gubin} \emph{et al.} provided a systematic study over general cyclic projections including the case of $C = \emptyset$. Based on these results, we have the following proposition.

\begin{proposition}
As $n$ increases, the optimal solution $\bm{x}_n^{(i)}$ of the $i$th BS's cyclic subproblem converges in norm to a limit $\bm{x}^i$ that lies in $F_i$. Moreover, if Problem (\ref{SOCP feasible}) is feasible, all $\bm{x}^i$'s coincide in a common point $\bm{x}^*$ that lies in $F$. If Problem (\ref{SOCP feasible}) is infeasible, $\bm{x}^i$'s do not coincide in the same solution.
\end{proposition}

\emph{Proof}: It is obvious that the optimal solutions for cyclic subproblems in (\ref{subproblem cyclic}) form a sequence of cyclic projections. Since weak convergence is always guaranteed \cite{Bregman}, by the equivalence of weak convergence and convergence in norm in a finite dimensional space \cite{Deutsch1}, we obtain Proposition 2.~$\Box$

\begin{rem}
Note that the convergence proof of CPB is more general than that of APB since alternating projections is actually a special case of cyclic projections where the number of projection sets is two.
\end{rem}

\subsection{Practical Feasibility Decision Rules}
For CPB, the algorithm termination rules are similar to that of APB, which is skipped here.

\section{Simulation Results}\label{simulations}

\begin{figure}[htbp]
    \begin{center}
    \includegraphics[width=3.0in]{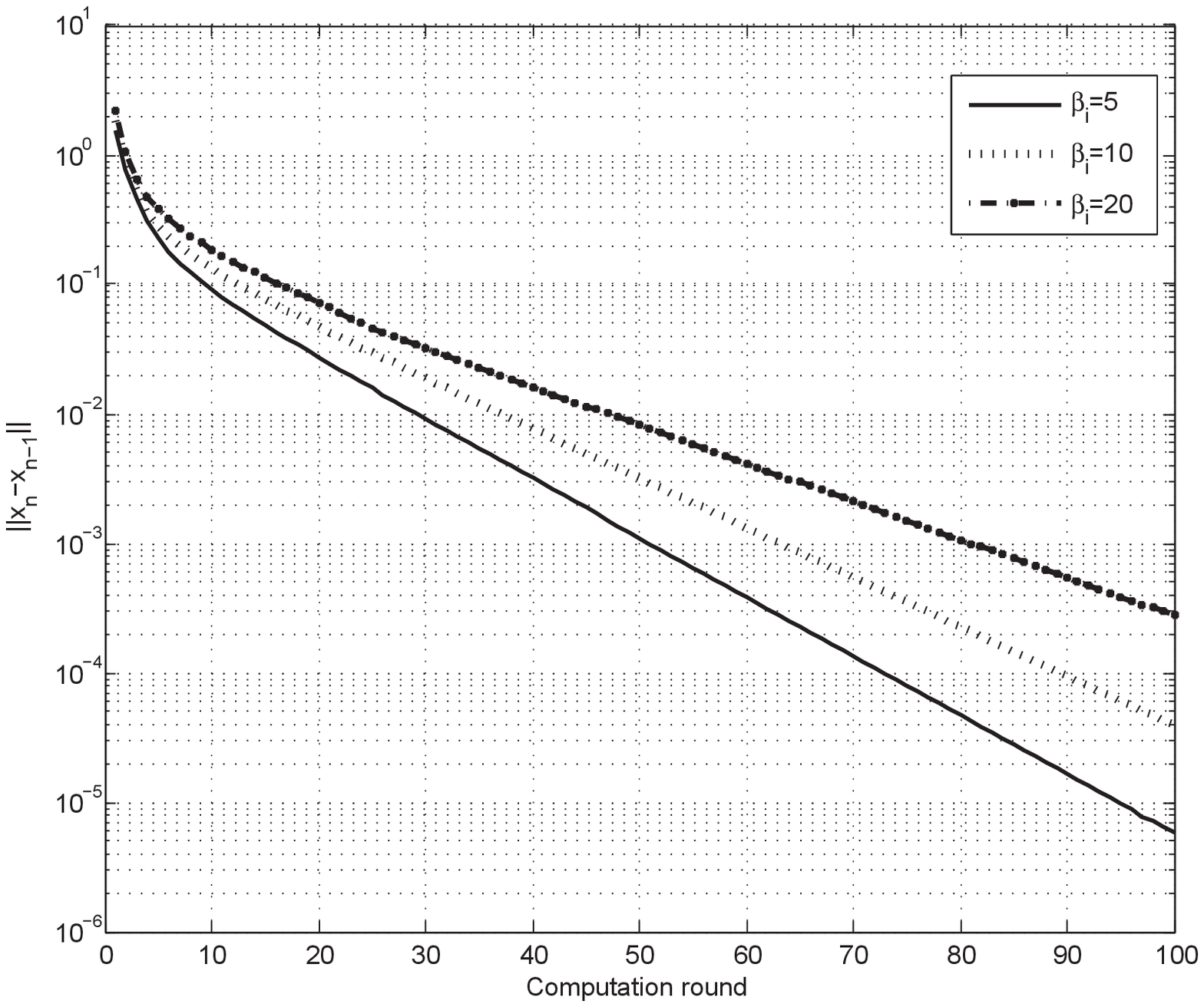}
    \caption{APB: $\left\| {{{\widetilde x}_n} - {{\widetilde x}_{n - 1}}} \right\|$ decreases}
    \label{fig 1}
    \end{center}
\end{figure}

\begin{figure}[htbp]
    \begin{center}
    \includegraphics[width=3.0in]{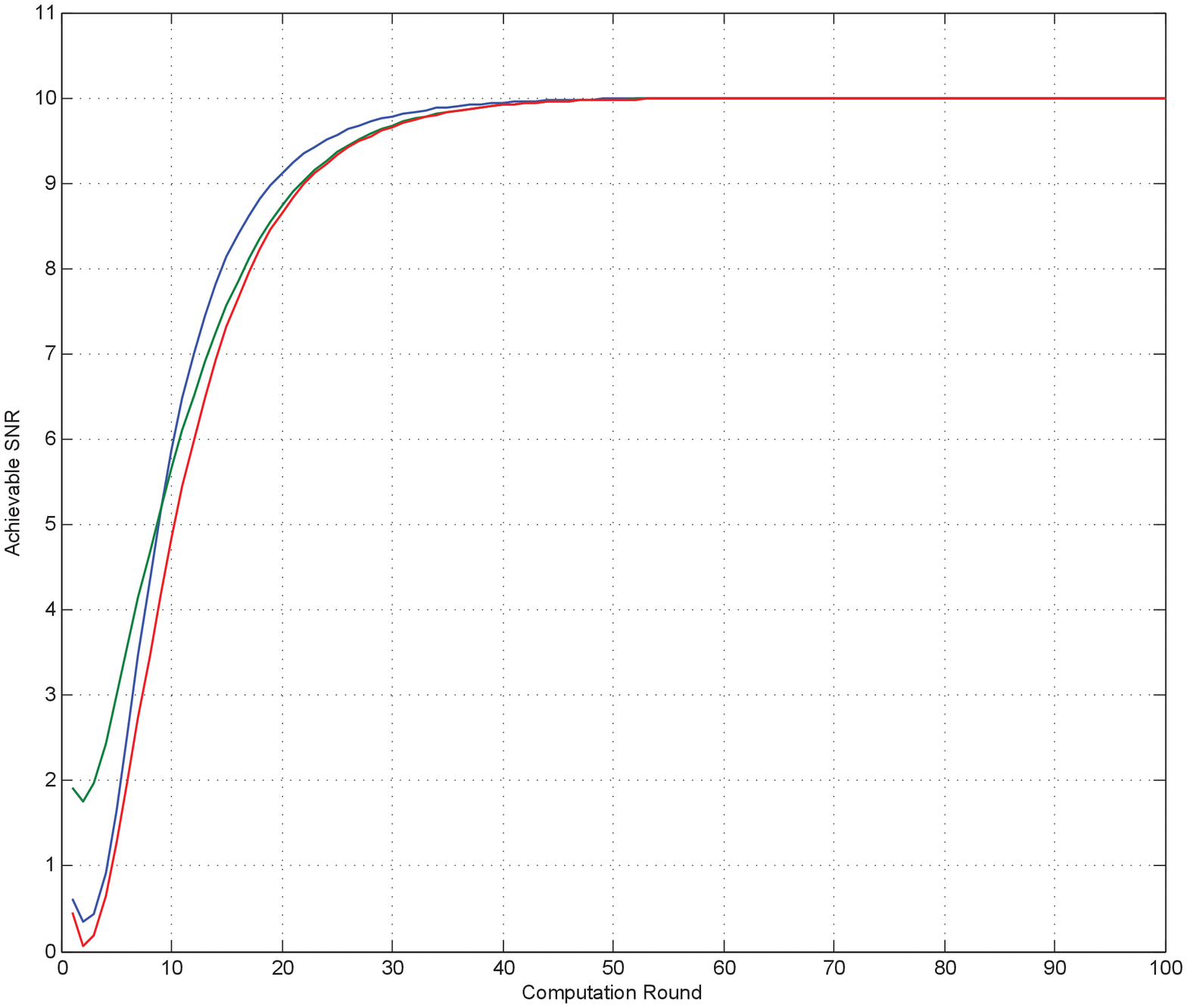}
    \caption{APB: Achievable SNR tuple increases, on setting $[10~10~10]$}
    \label{fig 3}
    \end{center}
\end{figure}

\begin{figure}[htbp]
    \begin{center}
    \includegraphics[width=3.0in]{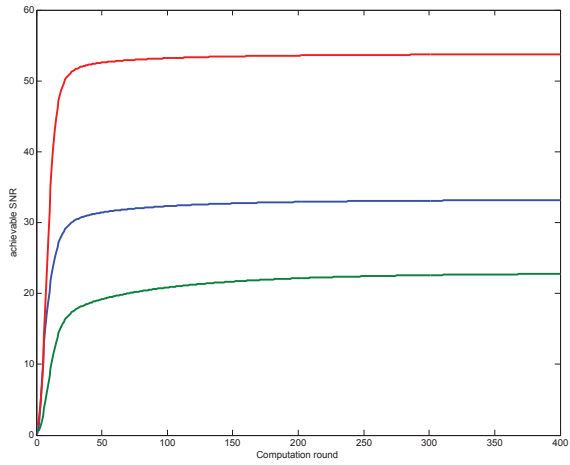}
    \caption{APB: Achievable SNR tuple increases, on setting $[50~40~60]$}
    \label{fig 5}
    \end{center}
\end{figure}

The performance of APB is first simulated. In the simulations, we set $M = 3$ and $K=4$. We set the power constraints as 15, 18, and 21, respectively, for the three BSs. In Fig.~\ref{fig 1}, we demonstrates the convergence behavior as described in Proposition 1. The three curves correspond to the required SNR ${\beta}_i$'s as 5, 10, and 20, respectively. We observe that their asymptotic behaviors are similar. In Fig. \ref{fig 3}, with a feasible choice of ${\beta}_i = 10$, $i = 1, 2, 3$, we draw how the  achieved SNR values approach the target values over iterations. If Problem (\ref{SOCP feasible}) is infeasible, for example, when setting target SNR as $[50~40~60]$, the SNR evolution curves are given in Fig. \ref{fig 5}, where we see that none of the target SNRs are satisfied.

\begin{figure}[htbp]
    \begin{center}
    \includegraphics[width=3.0in]{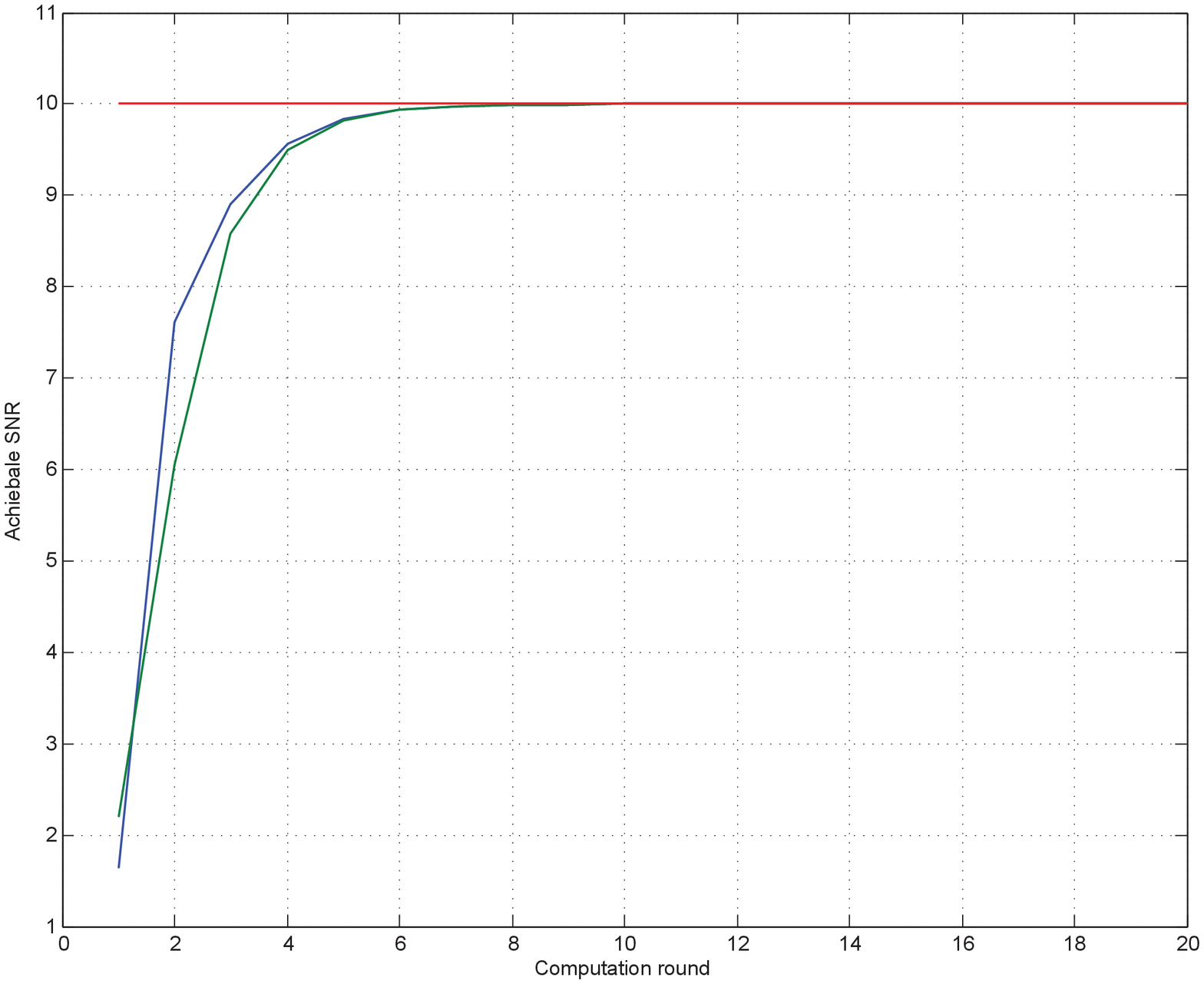}
    \caption{CPB: Achievable SNR tuple increases, on setting $[10~10~10]$}
    \label{fig 8}
    \end{center}
\end{figure}

For the performance of CPB, the simulation setup is exactly the same as that for APB. In Fig. \ref{fig 8}, with a feasible choice of ${\beta}_i = 10$, $i = 1, 2, 3$, we see how the achieved SNR values approach the target values with less iterations needed compared with Fig. \ref{fig 3}.


We observe from multiple simulations that the convergence speed of CPB is much faster than APB.

\section{Concluding Remarks}\label{conclusion}
In this paper, based on alternating projections and cyclic projections, we have developed two optimal distributed beamforming schemes to cooperatively solve the SOCP feasibility problem that is the key for quantifying the Pareto optimal points in the achievable rate region of MISO interference channels. The convergence in norm for both algorithms was established, which was further verified by numerical simulations.

\end{document}